\documentclass[usenatbib]{mn2e}
\usepackage{graphicx}
\usepackage{amssymb}
\usepackage{times}

\begin{document}

\title[Galaxy Interactions and the Mass-Metallicity Relation]
{The mass-metallicity relation of interacting galaxies}

\author[Michel-Dansac et al.]{L. Michel-Dansac,$^{1,2}$\thanks{E-mail: leo@oac.uncor.edu},
  Diego G. Lambas, $^{1,2}$,
  M. Sol Alonso, $^{1,3}$ and
  Patricia Tissera, $^{1,4}$\\
  $^{1}$ Consejo Nacional de Investigaciones Cient\'{\i}ficas y T\'ecnicas.\\
  $^{2}$ IATE, CONICET, OAC, Universidad Nacional de C\'ordoba,  Argentina.\\
  $^{3}$ Complejo Astron\'omico El Leoncito, Argentina \\
  $^{4}$ Instituto de Astronom\'{\i}a y F\'{\i}sica del Espacio, Argentina.\\
}

\maketitle

\begin{abstract}

  We study the mass-metallicity relation of galaxies in pairs and in isolation
  taken from the SDSS-DR4 using the stellar masses and oxygen abundances
  derived by Tremonti et al. (2004).  Close galaxy pairs, defined by projected
  separation $r_{p} < 25$~kpc~$h^{-1}$ and radial velocity $\Delta V < 350$
  ${\rm km\ s^{-1}}$, are morphologically classified according to the strength
  of the interaction signs.  We find that only for pairs showing signs of
  strong interactions, the mass-metallicity relation differs significantly
  from that of galaxies in isolation.  In such pairs, the mean gas-phase
  oxygen abundances of galaxies with low stellar masses (M$_\ast \la
  10^9$M$_\odot h^{-1}$) exhibit an excess of 0.2 dex. Conversely, at larger
  masses (M$_\ast \ga 10^{10}$M$_\odot h^{-1}$) galaxies have a systematically
  lower metallicity, although with a smaller difference ($-0.05$ dex). Similar
  trends are obtained if g-band magnitudes are used instead of stellar masses.
  In minor interactions, we find that the less massive member is
  systematically enriched, while a galaxy in interaction with a comparable
  stellar mass companion shows a metallicity decrement with respect to
  galaxies in isolation.

  We argue that metal-rich starbursts triggered by a more massive component,
  and inflows of low metallicity gas induced by comparable or less massive
  companion galaxies, provide a natural scenario to explain our findings.

\end{abstract}


\section{INTRODUCTION}

\begin{figure*}
\includegraphics[width=88mm]{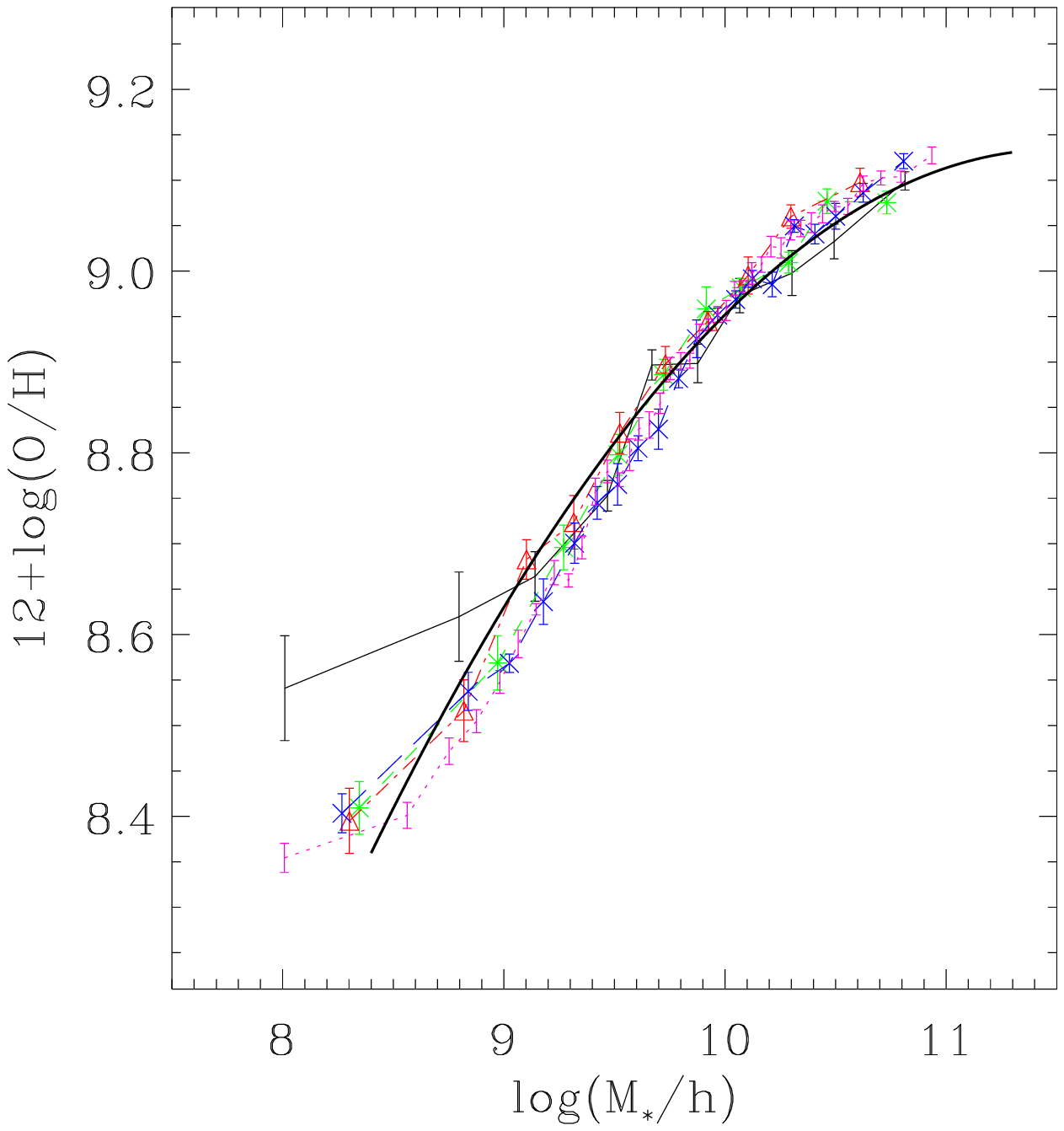}
\includegraphics[width=88mm]{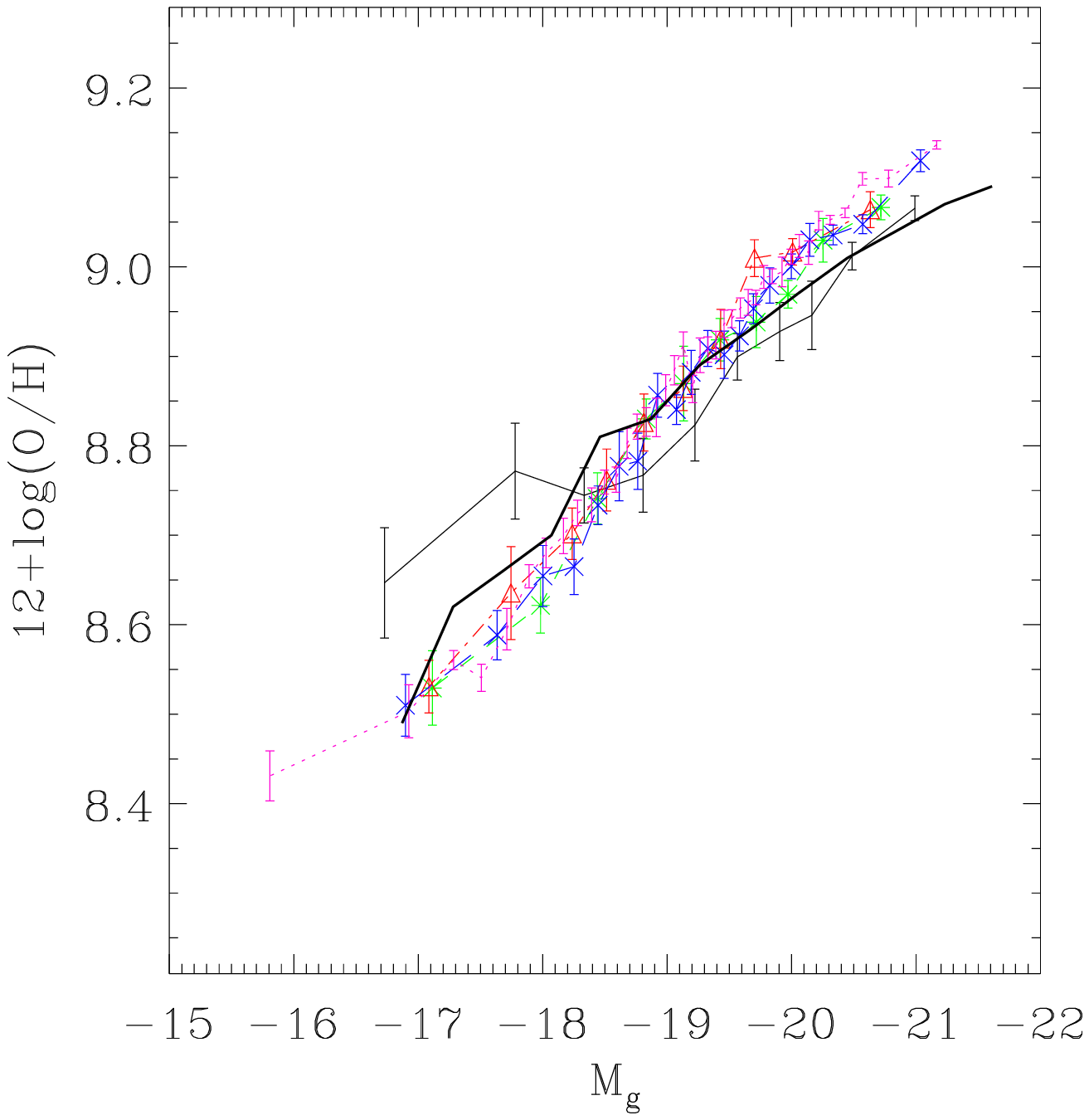}
\caption{Mass-metallicity (left) and Luminosity-metallicity (right) for
  galaxies in M (black solid lines), T (green dashed lines), N (red
  dotted-dashed lines), and D (blue long-dashed lines) type pairs, and in the
  control sample (magenta dotted lines). The thick solid line corresponds to
  Tremonti et al mean values.  Error bars have been estimated by applying the
  bootstrap resampling technique.  }
\label{meanmzr}
\end{figure*}

Chemical features observed in galaxies can store fossil records of their
history of formation since they are the result of different physical
mechanisms acting at different stages of evolution \citep{FBH02}.  Among them,
galaxy interactions and mergers are considered important processes which can
affect the star formation activity significantly as has been reported by
numerous works (e.g. \citealt{SP67,TL78,BGK00,LTAC03,NCA04,ALTC06}). The link
of interactions and mergers with chemical properties have been recently
started to be addressed (\citealt{DP00,MMMVBG02,Fetal04,KGB06}).  Recently,
\citet{KGB06} found the metallicity in the nuclear regions of galaxies in very
close pairs to be displaced to lower levels than those measured in more
distant pairs and isolated galaxies. The authors claimed this effect to be
caused by gas inflows triggering by the interactions.

In this letter, we study the mass-metallicity relation (MZR,
\citealt{Letal79}) of close galaxy pairs as a tool to study the effects of
interactions on metallicity. In the local Universe, \citet{Tetal04} have
confirmed the dependence of metallicity on stellar mass with high stastistical
signal. \citet{Eetal06} has extended the study to high redshift finding a
similar correlation, although displaced to lower metallicity (see also
\citealt{Metal07}).  The luminosity-metallicity relation (LMR) has been also
well estimated in local Universe when stellar masses are not possible to
measure. However, it has been shown that luminosity is a less fundamental
parameter than mass (e.g. \citealt{Eetal06}). Despite these new results, the
origin of the MZR and its evolution remain unclear.  Mergers and interactions
could play an important role in shaping the MZR.  Hence, a statistical
analysis of the effects of interactions on chemical properties could help to
shed light on this issue.

Numerical simulations showed how tidal interactions can trigger star formation
activity and can modify the morphology of galaxies (e.g.  \citealt{Ketal98}
and references therein; \citealt{MH96,B04}). Gas inflows generated during
these interactions feed star formation activity (e.g. \citealt{BH96,T00}).
\citet*{TdRS05} showed that simulated galaxies in the concordance
$\Lambda$-CDM cosmology reproduce the shape of the MZR but not its evolution.
Part of the failure could be caused by the lack of Supernova (SN) feedback in
these simulations.  In fact, \citet{Betal07}, \citet{Getal07}, and \citet{FD07},
resorting to ad-hoc parametrizations for SN feedback model, claimed to be able
to reproduce the observed MZR in a cosmological context.

We performed a statistical analysis of the chemical properties of galaxies in
pairs selected from the SDSS-DR4, focusing on the MZR. Since it is already
clear that proximity is the main parameter associated to important increases
in the star formation activity, hereafter, we will centre our analysis on
close galaxy pairs.  These pairs have been reclassified according to the level
of morphological disturbances detected by inspection of their images in
merging, tidal and non-perturbed \citep{ALTC07}. Merging pairs are those
exhibiting the higher level of star formation activity.

This letter is organized as follows. In Section 2 we described the galaxy
samples. Section 3 discussed the results. The conclusions summarize the main
findings.

\section{DATA}

The main SDSS galaxy sample is essentially a magnitude limited spectroscopic
sample \textit{$r_{lim}$}$ < 17.77$ with most of the galaxies spanning a
redshift range $0 < z < 0.25$ with a median redshift at 0.1 \citep{Setal02}.
Following \citet[][and references therein]{ALTC06}, we selected galaxy pairs
by requiring members to have a relative projected separation $r_{\rm
  p} < 100$~kpc~$h^{-1}$ and a relative radial velocity $\Delta V <
350$~km~s$^{-1}$.

In order to properly assess the significance of the results obtained from the
pair catalogue, we defined a control sample by a Monte Carlo algorithm that
selects for each galaxy in a pair, two galaxies with similar redshift and
r-band luminosity but without a close companion within the adopted relative
velocity and separation thresholds.  This procedure assures that the control
catalogue has the same selection effects than the pair sample, and
consequently, it can be used to estimate the actual difference between
galaxies in pairs and in isolation.

\citet{ALTC06,ALTC07} discussed possible effects of aperture and
incompleteness, which were found to have no significant consequences in the
analysis.  Active galactic nuclei (AGN) could contribute to the emission
spectral features affecting our interpretation. Hence, AGNs had not been
included in neither the control nor the pair catalogues.

As mentioned above, we focus our analysis on the close galaxy pairs defined as
those with a relative projected separation $r_{\rm p} < 25$~kpc~$h^{-1}$.
These pairs are subclassified accordingly to the level of morphological
disturbances in non-disturbed (N-type), tidal (T-type) and merging (M-type)
pairs as discussed by \citet{ALTC07}.  M-type pairs show clear and strong
signals of interactions and have active star formation.  For completeness, we
have also included the estimations for more distant pairs, $50$~kpc~$h^{-1} <
r_{\rm p} < 100$~kpc~$h^{-1}$ (D-type).

For the SSDS-DR4, \citet{Tetal04} estimated gas-phase oxygen abundance,
$12+$log O/H, and derived stellar masses ($M^*$).  In Table ~\ref{table1}, we
show the numbers of pairs and galaxies in the different samples which have
oxygen abundances measured for both members or for only one member.  Pairs
with only one member with oxygen estimations have been used to improve the
statistics, when possible.

\begin{table}
  \caption{Number of pairs and galaxies in the different samples 
    used in this work ($N_p$ total number of pairs; $N_p^1$ number of pairs
    with oxygen  abundances measured for both members;
    $N_g$ number of galaxies with oxygen abundances estimated for at
    least one member)}
  \begin{tabular}{@{}lccccc}
    \hline
    Sample  &  M  &  T  &  N  &  D  & C \\
    \hline
    $N_p$   & 383 & 686 & 536 & 4982 & - \\
    $N_p^1$ &  94 & 142 & 124 & 679 & - \\
    $N_g$   & 284 & 454 & 410 & 1358 & 6034 \\
    \hline
  \end{tabular}
  \label{table1}
\end{table}

\section{ANALYSIS AND RESULTS}

In Fig.~\ref{meanmzr}, we show the mean MZR and LZR for galaxies in M, T, N
and D-type pairs and the control sample. We also plot the fit to the observed
MZR given in Eq.~3 by \citet{Tetal04}, and the mean value of the observed LZR
(also given by \citealt{Tetal04}). It can be appreciated that our different
samples are in general agreement with these mean relations except for M-type
pairs. For this sample, at low stellar mass, or low luminosity, there is a
significant excess in the gas-phase metallicity with respect to the other
samples and mean observed values of \citet{Tetal04}.  Conversely, at high mass
(or high luminosity), we find M-type pairs galaxies to have slightly smaller
metallicity. This behaviour is in agreement with the results reported by
\citet{KGB06}. These authors found a decrease in the metallicity in the
nuclear region of galaxies in close pairs which had members with luminosities
consistent with our high luminosity end. 

We have also explored the dependence of the MZR on galaxy concentration for
the samples analised finding a similar behaviour irrespective of galaxy
concentration (here concentration is defined as the ratio of Petrosian 90--50
per cent r-band light radii).  Since the concentration parameter is well
correlated to morphology (see however \citet{ellis} for a detailled discussion
of the relation of morphology and other galaxy properties), this result
indicates that our findings are not merely a morphology-induced effect.

The change in the slope of the MZR of M-type pairs as a function of $M^*$
shown in Fig.~\ref{meanmzr} suggests a possible dependence of the effects of
interactions on the stellar mass of the members.  Hence, we splitted the pair
samples into minor and major interaction subsamples by adopting a ratio of
0.20 between the corresponding stellar masses.  We studied the MZR and star
formation activity for each component in major and minor interactions in the
four defined pair samples.

\begin{figure}
\resizebox{\hsize}{!}{\includegraphics{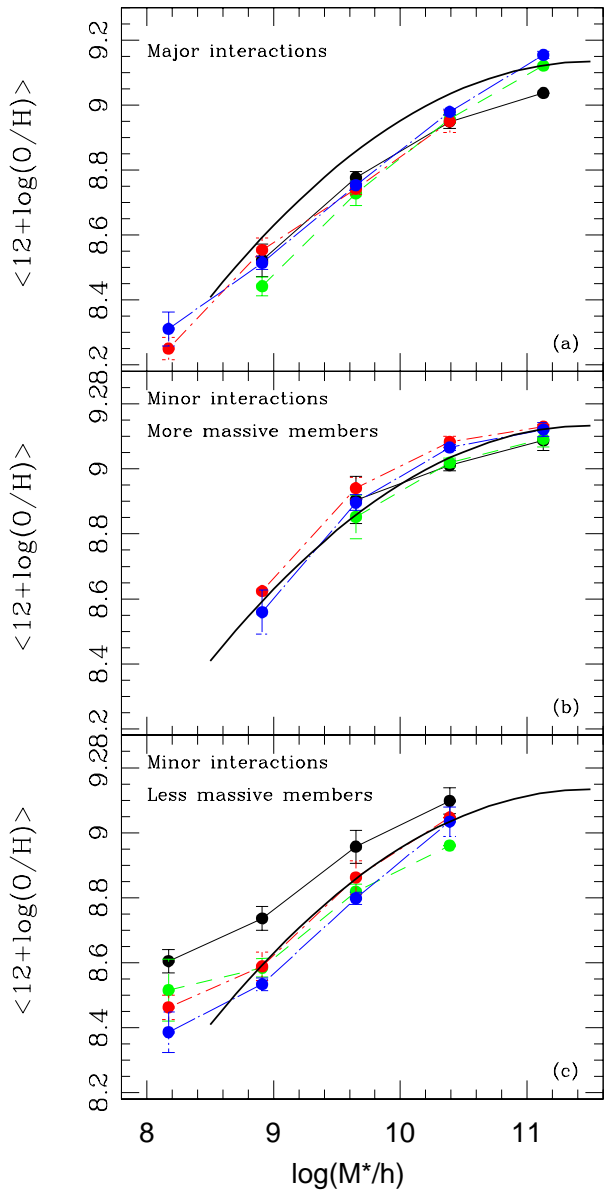}}
\caption{Mass - metallicity relation for the different type pairs: M (black), T
(green), N (red), and D (blue);
in major interactions ($M^*_2/M^*_1>5$) (a) and in minor interactions
($M^*_2/M^*_1<5$) (b: massive components; c: less massive components).
The thick solid line corresponds to Tremonti et al fit.
Error bars have been estimated by applying the bootstrap resampling technique.
}
\label{minormajor}
\end{figure}

As shown in Fig.~\ref{minormajor}, two cases depart considerably from the
global relation given by \citet{Tetal04}. Galaxies in major interactions
(Fig.~\ref{minormajor}a), which show a lower metallicity, and the small
component of minor interactions (Fig.~\ref{minormajor}c) with strong
disturbances (M-type) which exhibit a significantly larger mean metallicity at
a given stellar mass.

These results suggest that in major interactions, there is an efficient inflow
of pristine gas from the external regions of the halo. We argue that the lack
of an effect for the large component ($M^*_2$) in minor interactions may
reside on the relatively weak perturbation these galaxies are subject to. On
the contrary the small component ($M^*_1$) of minor interactions are the ones
mostly disturbed by the interactions and therefore those subject to major
episodes of star formation and metallicity enhancement.

\begin{figure}
\resizebox{\hsize}{!}{\includegraphics{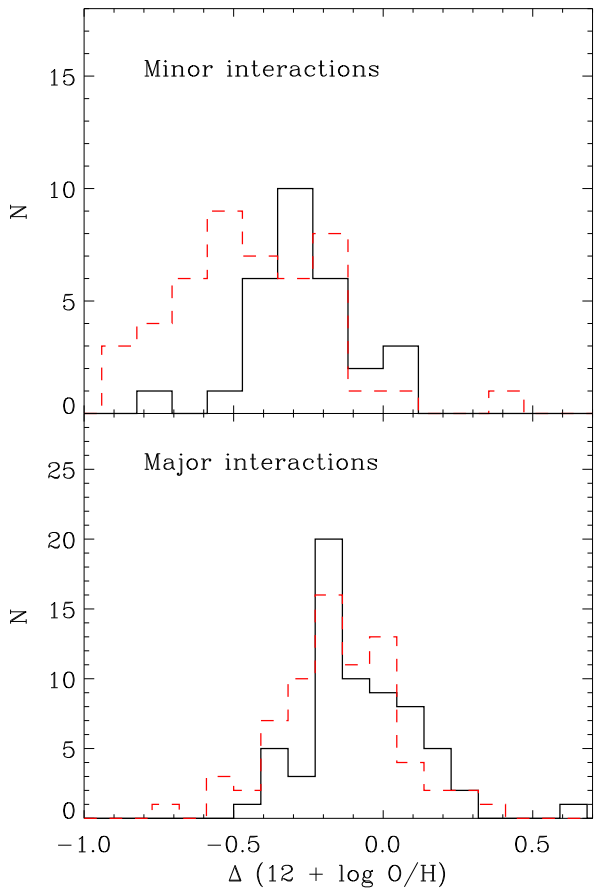}}
\caption{Distribution of oxygen abundance differences between the less massive
  and the most massive components in M-type (black solid lines) and N-type
  (red dashed lines) pairs in minor (upper panel) and major (lower panel)
  interactions.  }
\label{hist}
\end{figure}

We argue that the ratio between the stellar masses of the galaxies involved in
an interaction is an important parameter in setting the metallicity content of
galaxies. To explore this, we show in Fig.~\ref{hist} (upper panel) that the
metallicity difference ($\Delta$) between the less massive and the most
massive components in M-type minor interactions is significantly reduced in
comparison to that of galaxies in N-type pairs (similar to D-type pairs).
Members of major interactions in M-type pairs have similar metallicity
differences than those of N-type pairs (lower panel in Fig.~\ref{hist}).

An analysis of the star formation activity indicates that metal-rich less
massive members in M-type pairs are in general also those exhibiting a high
star formation activity, at a given stellar mass. This somewhat expected
result reinforces the idea that minor components in strong interactions are
likely to have experienced strong gas shocks leading to star formation and
enrichment.

\begin{figure}
\resizebox{\hsize}{!}{\includegraphics{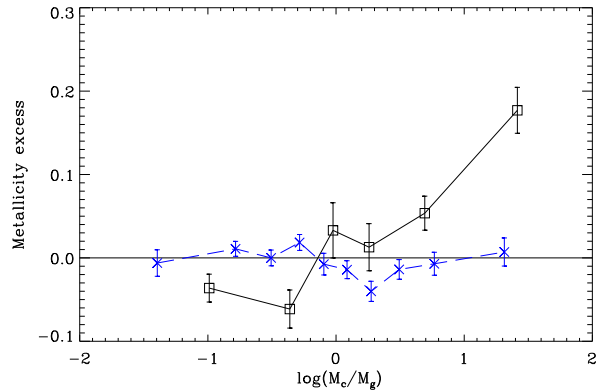}}
\caption{ Metallicity excess of a given galaxy with respect to galaxies in the
  control sample with similar stellar mass as a function of mass ratio between
  the companion galaxy and the given one, for M (solid line), and D
  (long-dashed line) type pairs.  Error bars correspond to the mean standard
  deviation.  }
\label{dglplot}
\end{figure}

In order to further assess the role of the relative mass ratio of the galaxy
members on the mean gas-phase metallicity, we calculated the metallicity
excess ($\epsilon$) as the difference between the oxygen abundance of a given
galaxy in a pair and the mean abundance of galaxies in the control sample with
similar stellar mass as a function of the ratio between the stellar mass of
the companion ($M_{c}$) galaxy and the stellar mass of the given galaxy
($M_{g}$).  We estimated $\epsilon $ for all subsamples. In Fig.~\ref{dglplot},
we display the results corresponding to pairs in the M-type and D-type
subsamples since these galaxies set the range of variation. As it can be
appreciated, in M-type pairs, galaxies with massive companion (log
$M_{c}/M_{g} > 0 $) have significantly enhanced their metallicity with respect
to galaxies in the control sample with the same stellar mass. The opposite
occurs for galaxies with a less massive companion (log $M_{c}/M_{g} < 0 $)
where the metallicity is lower.  Conversely, D-type pairs exhibit a remarkable
consistency with the control sample.  The behaviour of M-type pairs stresses
that the relative mass ratio plays a major role in determining the metallicity
properties in galaxies in strong interactions.

\section{DISCUSSION AND CONCLUSIONS}

We analyse the effects of interaction on the metallicity properties of
galaxies. For this aim we have studied pairs of galaxies with projected
separation $r_p < 25$~kpc~$h^{-1}$ and radial velocity $\Delta V < 350 \ {\rm
  km^{-s}}$ from the SDSS-DR4 and oxygen abundances from \citet{Tetal04}.
Using the visual classification of pairs into M-type, T-type, N-type, and
D-type pairs of Alonso et al. (2007), we find that pairs undergoing a strong
interaction differ significantly on the mass metallicity relation in
comparison to other pair categories and a control sample.  In the low mass
range (M$_\ast \la 10^9$M$_\odot h^{-1}$) the mean metallicity of galaxies in
M-type pairs is 0.2 dex greater. On the contrary, at large masses (M$_\ast \ga
10^{10}$M$_\odot h^{-1}$), M-type pair galaxies have a systematically lower
metallicity, although with a smaller difference ($- 0.05$ dex).

When dividing the pairs into minor (M$^*_2/$M$^*_1>5$) and major
(M$^*_2/$M$^*_1<5$) interactions, we find that the less massive member in
strong minor interactions are always enriched irrespective of their mass.  The
interactions of galaxies of comparable stellar mass content have a
systematically lower metallicity content, suggestive of an inflow of
unenriched material from the external regions.

We notice that the finite size of the fiber might affect the results in the
sense that the smaller the angular size of the objects the larger the region
where metallicity is derived. Nevertheless, in this work we have considered
galaxies of similar stellar masses and redshifts differing only on the
presence of a companion.  In order to analyse this effect on the MZR into more
detail, we have considered separately large and small angular size galaxies
(Petrosian 90 per cent r-band light radius $\gtrless 8$~arcsec) for the
samples in Fig.~\ref{meanmzr}. We find that the trends obtained in this work
remain unchanged regardless of the apparent angular size of the galaxies.
Therefore, the results are robust against a combination of observational
aperture and metallicity gradient in the galaxies.

We may interpret our results in terms of a combination of the effects of a
metal-rich starburst induced by a massive component and the pristine gas
inflow triggered by comparable or less massive companion galaxies. These two
effects are expected naturally in current galaxy formation and evolution
scenarios, so our findings may help to understand the origin and evolution of
the mass-metalicity relation.

\section*{Acknowledgments}

We thank the Referee for helpful comments and suggestions.
This work was partially supported by the
 Consejo Nacional de Investigaciones Cient\'{\i}ficas y T\'ecnicas,
Agencia de Promoci\'on de Ciencia y Tecnolog\'{\i}a,  
 and Secretar\'{\i}a de Ciencia y
T\'ecnica de la Universidad Nacional de C\'ordoba.

\end{document}